\documentclass[a4paper,UKenglish,cleveref, autoref, thm-restate]{lipics-v2021}
\usepackage{amsmath,amsthm,amssymb, color, tikz, xcolor}
\usepackage{tcolorbox}
\usepackage{thmtools} 
\usepackage{bbm}
\usepackage{dsfont}

\usetikzlibrary{positioning}
\usetikzlibrary{shapes}
\usetikzlibrary{decorations.pathreplacing}
\usepackage{enumerate} 
\usepackage{bookmark}
\usepackage{algorithm}
\usepackage{algpseudocode}

\usepackage{mathtools}
\makeatletter
\newenvironment{proof*}[1][\proofname]{\par
  \pushQED{\qed}%
  \normalfont \partopsep=\z@skip \topsep=\z@skip
  \trivlist
  \item[\hskip\labelsep
        \itshape
    #1\@addpunct{.}]\ignorespaces
}{%
  \popQED\endtrivlist\@endpefalse
}
\makeatother

\newcommand{\mathify}[1]{\ifmmode{#1}\else\mbox{$#1$}\fi}

\newcommand{\ignore}[1]{}
\newcommand{\stream}{\mathcal{A}}

\newcommand{\hybrid}{\ensuremath{\mathsf{F_0\mbox{-}Estimator}}}
\newcommand{\Fo}{\ensuremath{\mathsf{F_0}}}

\newcommand{\p}{\mathsf{p}}

\newcommand{\Fail}{\mathsf{Fail}}
\newcommand{\Error}{\mathsf{Error}}
\newcommand{\bad}{\ensuremath{\mathsf{Bad}_{2}}}
\newcommand{\Bad}[1]{\ensuremath{\mathsf{Bad}_{2,{#1}}}}

\newcommand{\thresh}{\mathsf{thresh}}

\newcommand{\poly}{\ensuremath{\mathsf{poly}}}

 \newtheorem{fact}{Fact}[section]
 \newtheorem{problem}[theorem]{Problem}
\usepackage{bbm} %
\usepackage{mathtools}

\algnotext{EndFor}
\algnotext{EndIf}
\algnotext{EndWhile}
\algblockdefx{Do}{EndDo}{\textbf{do}\;}{}
\algnotext{EndDo}
\algnewcommand\algorithmicswitch{\textbf{switch}}
\algnewcommand\algorithmiccase{\textbf{case}}
\algdef{SE}[SWITCH]{Switch}{EndSwitch}[1]{\algorithmicswitch\ #1\ \algorithmicdo}{\algorithmicend\ \algorithmicswitch}%
\algdef{SE}[CASE]{Case}{EndCase}[1]{\algorithmiccase\ #1}{\algorithmicend\ \algorithmiccase}%
\algtext*{EndSwitch}%
\algtext*{EndCase}%

\usepackage[colorinlistoftodos,prependcaption,textsize=tiny]{todonotes}

\title{Distinct Elements in Streams: An Algorithm for the (Text) Book\thanks{The earlier version of the paper, as published at ESA 2022, contained error in the proof of  Claim~\ref{claim:errfail}. The current revised version fixes the error in the proof. The authors decided to forgo the old convention of alphabetical ordering of authors in favor of a randomized ordering, denoted by \textcircled{r}. The publicly verifiable record of the randomization is available at \protect\url{https://www.aeaweb.org/journals/policies/random-author-order/search}}} %

\author{Sourav Chakraborty \textcircled{r}}{Indian Statistical Institute, India %\and 
}{}{}{}%

\author{N.~V.~Vinodchandran \textcircled{r}}{University of Nebraska-Linclon, USA %\and 
}{}{}{}

\author{Kuldeep~S.~Meel}{National University of Singapore, Singapore 
}
{}{}{}

\authorrunning{S. Chakraborty, N. V. Vinodchandran and K. S. Meel} %
\titlerunning{Distinct Elements in Streams: An Algorithm for the (Text) Book}
\Copyright{S. Chakraborty, N. V. Vinodchandran and K. S. Meel} %



\category{} %

\nolinenumbers %

\begin{document}

\maketitle
\begin{abstract}
 Given a data stream $\stream = \langle a_1, a_2, \ldots, a_m \rangle$ of $m$ elements where each $a_i \in [n]$, the Distinct Elements problem is to estimate the number of distinct elements in $\stream$. 
 Distinct Elements has been a subject of  theoretical and empirical investigations over the past four decades resulting in space optimal algorithms for it. 
 All the current state-of-the-art algorithms are, however, beyond the reach of an undergraduate textbook owing to their reliance on the usage of notions such as pairwise independence and universal hash functions. We present a simple, intuitive, sampling-based space-efficient algorithm whose description and the proof are accessible to undergraduates with the knowledge of basic probability theory. 
\end{abstract}
%

%

%

%\section{A Simple Algorithm for Distinct Elements Estimation}
\section{Introduction}
We consider the fundamental problem of estimating the number of distinct elements  in a data stream (Distinct Elements problem or the $\Fo$ estimation problem). For a data stream $\stream = \langle a_1,a_2,\ldots, a_m\rangle$, where each $a_i\in [n]$, $\Fo(\stream)$ is the number of distinct elements in $\stream$: $\Fo(\stream) = |\{a_1,a_2,\ldots,a_m\}|$. Since $\stream$ is clear from the context, we use $\Fo$ to refer to $\Fo(\stream)$. 

\begin{problem}\label{prob:delphic}
 Given a stream $\stream= \langle a_1, a_2, \ldots, a_m\rangle$ of $m$ elements where each $a_i \in [n]$, parameters $\varepsilon,\delta$, output an $(\varepsilon,\delta)$-approximation of $\Fo(\stream)$. That is, output 
 $c$ such that 
  $$\Pr[ (1-\varepsilon) \cdot \Fo(\stream) \leq c \leq (1+\varepsilon) \cdot \Fo (\stream)] \geq 1-\delta.$$
\end{problem}

 %$\Fo$ estimation problem is a fundamental problem with a long history of theoretical and practical investigations.
 We are interested in streaming algorithms that uses $\mathsf{poly}(\log m, \log n, \frac{1}{\varepsilon}, \log \frac{1}{\delta})$ bits of memory wherein we use $\log$ to denote $\log_2$. 
 The seminal work of Flajolet and Martin~\cite{FM85} provided the first algorithm assuming the existence of hash functions with full independence. Subsequent investigations relying on the usage of limited-independence hash functions have led to design of algorithms with optimal space complexity $O(\log n + \frac{1}{\varepsilon^2}\cdot \log\frac{1}{\delta})$. We defer detailed bibliographical remarks to Section~\ref{sec:related}. 
However, all the current space-efficient algorithms are beyond the reach of an undergraduate textbook due to their reliance on  notions such as pairwise independence and universal hash functions.

We present a very simple algorithm for the $\Fo$ estimation problem using  a sampling strategy  that only relies on basic probability for its analysis. In particular, it does not use  universal hash functions. We believe that only using basic probability theory for the analysis makes the algorithm presentable to undergraduates right after the introduction of basic tail bounds. In addition,  the simplicity of the code makes it appealing to be used in practical implementations.  Our algorithm builds and refines ideas introduced in the recent work on estimating the size of the union of sets in the general setting of {\em Delphic sets}~\cite{MVC21}.     

%

%
%
%

%

%

%

%
%
%
%
%
%
%
%

%

%
%
%
%

%

%
%
%
%

%
%

%

%

%

%

%
%
%
%

%
%
%
%
%
%
%
%

%
%
%
%

%
%
%

%

%
%
%\section{{\hybrid}: A simple algorithm for {\Fo}   estimation}\label{sec:hybrid}
\section{A Simple Algorithm}

\begin{algorithm}
\caption{$\hybrid$}\label{algo:afinal}
\begin{algorithmic}[1]
\Statex \textbf{Input} Stream $\stream = \langle a_1,a_2,\ldots,a_m \rangle$, $\varepsilon$, $\delta$
 \State \textbf{Initialize}  $p \gets 1$; $\mathcal{X} \gets \emptyset$; $\thresh \gets \lceil\frac{12}{\varepsilon^2}\log (\frac{8m}{\delta})\rceil$
 \For{$i = 1$ to $m$}\label{line:beginfor}
\State $\mathcal{X} \gets \mathcal{X} \setminus \{a_i\}$\label{line:afinal-drop}
\State With probability $p$, $\mathcal{X} \gets \mathcal{X} \cup \{a_i\}$ \label{line:afinal-pick}
\If{$|\mathcal{X}|= \thresh$}\label{line:Ifbegin}
\State Throw away each element of $\mathcal{X}$ with probability $\frac{1}{2}$ \label{line:final-throw}
\State $p \gets \frac{p}{2}$ \label{line:p}
\If{$|\mathcal{X}| = \thresh$}\label{line:check} \textbf{Output} $\bot$\label{line:fail}
\EndIf 
\EndIf \label{line:Ifend}
\EndFor \label{line:endfor}
\State \textbf{Output} $\frac{|\mathcal{X}|}{p}$\label{line:output}
\end{algorithmic}
\end{algorithm}

The algorithm $\hybrid$ uses a simple sampling strategy.
In order to keep the set of samples small, it makes sure that ${\mathcal{X}}$ does not grow beyond the value $\thresh$ by adjusting the sampling rate $p$ accordingly.
After all the elements of the stream are processed, it outputs
$\frac{|\mathcal{X}|}{p}$ where $p$ is the final sampling rate\footnote{In an earlier version, it was wrongly claimed that every element seen so far is independently in $\mathcal{X}$ with equal probability $p$. That claim was erroneous but, fortunately, not used in the analysis. It is also worth remarking that $\frac{|\mathcal{X}|}{p}$ is not an unbiased estimator of $\Fo$. }.

\subsection{Theoretical Analysis}
We present the theoretical analysis entirely based on first principles, which adds to its length. For readers who are familiar with randomized algorithms, the proof is standard. 

We state the following well-known concentration bound, Chernoff bound,  for completeness.

\begin{fact} [Chernoff's Bound] 
Let $\mathsf{v}_1, ..., \mathsf{v}_k$ be independent random variables taking values in $\{0, 1\}$. Let $\mathsf{V} = \sum_{i=1}^k \mathsf{v}_i$ and $\mu = \mathbb{E}[\mathsf{V}]$. Then, for $\beta > 0$,  
$\Pr\left(\left|\mathsf{V} - \mu\right| \geq \beta \mu\right) \leq 2e^{-\frac{\beta^2\mu}{2+\beta}}$
\end{fact}

The following theorem captures the correctness and space complexity of {\hybrid}.

\begin{theorem}
For any data stream ${\stream}$ and any $0< \varepsilon, \delta <1 $, the algorithm $\hybrid$ outputs an $(\varepsilon,\delta)$-approximation of $\Fo({\stream})$. The algorithm uses $O(\frac{1}{\varepsilon^2}\cdot\log n \cdot (\log m + \log \frac{1}{\delta}))$ space in the worst case. 
\end{theorem}

\begin{proof*}
The stated space complexity bound of the algorithm follows because, from the description, it is clear that the size of the set of samples kept by the algorithm is always $\leq \thresh$, and each item requires $\lceil\log_2 n\rceil$ bits to store.

We give a formal proof of correctness below. 
Consider the following two events: 

\begin{description}
\item[]$\mathsf{Error}:$ `The algorithm $\hybrid$ does not return a value in the range $[(1-\varepsilon)\Fo, (1+\varepsilon) \Fo]$'
\item[]$\mathsf{Fail}:$ `The algorithm $\hybrid$ outputs $\bot$.'
\end{description}

We will bound $\Pr[\mathsf{Error}]$ by $\delta$. Observe that $\Pr[\mathsf{Error}] \leq \Pr[\mathsf{Fail}] + \Pr [\mathsf{Error} \cap  \overline{\mathsf{Fail}}]$. Theorem follows from Claim~\ref{lm:fail} and Claim~\ref{claim:errfail}.
\end{proof*}
\begin{claim}\label{lm:fail}
$\Pr[\mathsf{Fail}] \leq \frac{\delta}{8}$
\end{claim}
\begin{proof*}
Let $\Fail_j$ denote the event that Algorithm~\ref{algo:afinal} returns $\bot$ when $i=j$. Formally, 
$\mathsf{Fail}_j$: `$|\mathcal{X}| = \thresh$ and none of the elements of $\mathcal{X}$ are thrown away at line~\ref{line:final-throw} for $i=j$'. 
 The probability that $\mathsf{Fail}_j$ happens is $\left(\frac{1}{2}\right)^{\thresh}$. 
Therefore,
 \begin{align*}
\Pr[\mathsf{Fail}] \leq \sum_{j=1}^{m} \Pr[\mathsf{Fail}_j] &\leq  m\cdot \left(\frac{1}{2}\right)^{\thresh} \leq \frac{\delta}{8}\qedhere 
 \end{align*}
\end{proof*}

\begin{claim}\label{claim:errfail}
$\Pr [\mathsf{Error} \cap \overline{\mathsf{Fail}}] \leq \frac{\delta}{2}$. 
\end{claim}
%We give a detailed proof of this claim below. 

\paragraph*{Proof of Claim~\ref{claim:errfail}}
To bound $\Pr [\mathsf{Error} \cap \overline{\mathsf{Fail}}]$, we consider a relaxed version of Algorithm~\hybrid, which is  stated as Algorithm~\ref{algo:bfinal}.  Algorithm~\ref{algo:bfinal} is nothing but \hybrid\ with line \ref{line:check} removed.
%and \ref{line:fail} removed. 
Observe that for a given input, the algorithm {\hybrid} behaves identically to Algorithm~\ref{algo:bfinal} as long as $|\mathcal{X}| < \thresh$ after each element of $\mathcal{X}$ is thrown away with probability $\frac{1}{2}$ (i.e., the event $\Fail$ does not happen). 
Now, we consider the following event:

\begin{description}
\item[]$\Error_{2}:$ `The Algorithm~\ref{algo:bfinal} does not output a value in the range $[(1-\varepsilon)\Fo, (1+\varepsilon) \Fo]$.'
\end{description}

Observe that $\Pr[\Error \cap \overline{\mathsf{Fail}}] \leq \Pr[\Error_{2}]$.
In Claim~\ref{lm:error-fail}, we obtain the desired bound on $\Pr[\Error_{2}]$ and hence on   $\Pr [\mathsf{Error} \cap \overline{\mathsf{Fail}}]$.

\begin{algorithm}
\caption{}\label{algo:bfinal}
\begin{algorithmic}[1]
\Statex \textbf{Input} Stream $\stream = \langle a_1,a_2,\ldots,a_m \rangle$, $\varepsilon$, $\delta$
 \State \textbf{Initialize}  $p \gets 1$; $\mathcal{X} \gets \emptyset$; $\thresh \gets \lceil\frac{12}{\varepsilon^2}\log (\frac{8m}{\delta})\rceil$
 \For{$i = 1$ to $m$}\label{line:bbeginfor}
\State $\mathcal{X} \gets \mathcal{X} \setminus \{a_i\}$\label{line:bfinal-drop}
\State With probability $p$, $\mathcal{X} \gets \mathcal{X} \cup \{a_i\}$ \label{line:bfinal-pick}
\If{$|\mathcal{X}|= \thresh$}\label{line:bIfbegin}
\State Throw away each element of $\mathcal{X}$ with probability $\frac{1}{2}$ \label{line:bhybrid-final-throw}
\State $p \gets \frac{p}{2}$ \label{line:bp}
\EndIf \label{line:bIfend}
\EndFor \label{line:bendfor}
\State \textbf{Output} $\frac{|\mathcal{X}|}{p}$\label{line:boutput}
\end{algorithmic}
\end{algorithm}

\begin{algorithm}
\caption{}\label{algo:cfinal}
\begin{algorithmic}[1]
\Statex \textbf{Input} Stream $\stream = \langle a_1,a_2,\ldots,a_m \rangle$ 
\For{$k=0$ to $m$}
 \State \textbf{Initialize}  $\mathcal{Y}_{k} \gets \emptyset$; %
\EndFor
 \For{$i = 1$ to $m$}\label{line:cbeginfor}
  \State $r \gets \mathsf{GenerateRandomBits}(m+1)$
 \For{$k = 0$ to $m$}
 \State $\mathcal{Y}_{k} \gets \mathcal{Y}_{k} \setminus \{a_i\}$\label{line:cfinal-drop}
 \If{$k \leq \mathsf{FirstZeroIndex}(r)$}  $\mathcal{Y}_{k} \gets \mathcal{Y}_{k} \cup \{a_i\}$ \label{line:cfinal-pick}
 \EndIf 
 \EndFor
\EndFor \label{line:cendfor}
% %
\end{algorithmic}
\end{algorithm}

%\subsubsection{Proof of}

To prove an upper bound on  $\Pr [\Error_{2}]$ in Claim~\ref{lm:error-fail}, we will consider the Algorithm~\ref{algo:cfinal}. Algorithm~\ref{algo:cfinal}  uses two subroutines: (1) $\mathsf{GenerateRandomBits}$ that returns a randomly generated array of $m+1$ bits, and (2) $\mathsf{FirstZeroIndex}$ returns the minimum of first index of array equal to $0$ and one plus the size of array;  we assume array is zero-indexed.
In the following, we use $S_i$ to denote $\{a_1,a_2,\ldots, a_i\}$ -- distinct elements that appear in the first $i$ items  in the stream.

\begin{claim}\label{cl:loop}
The following property holds true 
in line~\ref{line:cfinal-pick} of Algorithm~\ref{algo:cfinal}
%every iteration $i$ of the \textbf{for} loop (line~\ref{line:cbeginfor} to \ref{line:cfinal-pick} of Algorithm~\ref{algo:cfinal}),  for all $k$, in line~\ref{line:cfinal-pick} 
%in the \textbf{for} loop (lines~\ref{line:cbeginfor}-- \ref{line:cendfor}) : 
\[
\mbox{
Every element in $S_i$ is in $\mathcal{Y}_{k}$ independently with probability $2^{-k}$.
} 
\]
\end{claim}

\begin{proof*}%

The proof proceeds via induction on $i$. 

\medskip
\noindent\textbf{Base Case}: 
%Before any element of the stream arrives $\mathcal{Y}_k$ is empty and so the base case is vacuously true. 
$\Pr[a_1 \in \mathcal{Y}_{k}] = \Pr[k \leq \mathsf{FirstZeroIndex}(r)] = 1/2^k$.

\noindent\textbf{Inductive Step}:
%Let us assume by induction hypothesis, the desired invariant holds true after iteration $i = j-1$.
Note that, by induction hypothesis, for all $a_{\ell} \in S_i \setminus a_i$, we have $a_{\ell} \in  \mathcal{Y}_{k}$ independently with probability $1/2^k$. 
%Also note that whether $a_i \in \mathcal{Y}_{k}$ is independent of whether $a_{\ell}$ is in $\mathcal{Y}_k$ for all $a_{\ell} \in S_i \setminus a_i$.  
%Now for [a_i \in \mathcal{Y}_{k}]$, we first have  %there are two cases: (i)  $a_i \notin S_{j-1}$ and (ii) $a_i \in S_{j-1}$.
Now, for $a_i$, we have $a_i \notin \mathcal{Y}_{k}$ in line~\ref{line:cfinal-drop}.
%But in either case after line~\ref{line:cfinal-drop}, we have $\Pr[a_i \in \mathcal{Y}_{k}] =0$.
% \begin{itemize}
% \item[--]  $a_j \notin S_{j-1}$, then after the execution of line~\ref{line:cfinal-pick}, we have 
% $$\Pr [a_j \in \mathcal{Y}_{k}] = \Pr[k < \mathsf{FirstZeroIndex}(r)] = \frac{1}{2^k}.$$
% \item[--]  $a_j \in S_{j-1}$,  after line~\ref{line:cfinal-drop}, we have $\Pr[a_j \in \mathcal{Y}] =0$. After line~\ref{line:cfinal-pick}, we have $\Pr [a_j \in \mathcal{Y}_{k}] = 1/2^kp$. \qedhere 
% \end{itemize}
In line~\ref{line:cfinal-pick},
$\Pr [a_i \in \mathcal{Y}_{k}] = \Pr[k \leq \mathsf{FirstZeroIndex}(r)] = 1/2^k$  \qedhere

\end{proof*}

\

Let us use the random variables $\p_j$ and $\mathsf{X}_{j}$ to denote the value of $p$ and the set $\mathcal{X}$ at the end of the loop iteration with $i=j$ (in Algorithm~\ref{algo:bfinal}). Similarly, we will use the random variable $\mathsf{Y}_{k,j}$ to indicate the set $\mathcal{Y}_{k}$ at the end of loop iteration with $i=j$ (in Algorithm~\ref{algo:cfinal}). 

We can view Algorithm~\ref{algo:bfinal} as updating value of $p$ and $\mathcal{X}$ as the elements of stream $\stream$ are processed such that we have $(\p_j,\mathsf{X}_{j}) = (\p_j,\mathsf{Y}_{k,j})$ where $\p_j=2^{-k}$. It is perhaps worth observing that the value of $p$ in Algorithm~\ref{algo:bfinal} is always at least $2^{-m}$, which is why we have $k$ iterate over $[0,m]$ in Algorithm~\ref{algo:cfinal}.   
%We can view Algorithm~\ref{algo:bfinal} inducing distribution over $(\mathcal{X}, p)$ such that $\mathcal{X} = \mathcal{Y}_{j,q} \wedge p= q$ after processing first $j$ elements.

%

%
For any $j \in [1, m]$ and $k\in [0,m]$ and $a\in S_j$ let $\mathsf{r}^a_{k,j}$ denote the indicator random variable indicating whether $a$ is in the set $\mathcal{Y}_{k}$ in line~\ref{line:cfinal-pick} for $i = j$ (in Algorithm~\ref{algo:cfinal}).  By Claim~\ref{cl:loop}, the random variables $\{\mathsf{r}^a_{k, j}\}_{a\in S_j}$ are independent and for all $a\in S_j$ $\Pr[\mathsf{r}^a_{k,j} = 1] = 2^{-k}$. 
%
%Since, $|\mathsf{Y}_{k,j}| = \sum_{a\in S_{j}} \mathsf{r}^a_{k, j}$
\begin{equation}\label{eq:Yellj}\mathbb{E}\left[|\mathsf{Y}_{k,j}|\right] = \mathbb{E}\left[\sum_{a\in S_{j}}\mathsf{r}^a_{k,j}\right] = \sum_{a\in S_{j}}\Pr[\mathsf{r}^a_{k,j} = 1] = 2^{-k}\cdot |S_{j}| \leq  2^{-k} \cdot \Fo 
%\leq \frac{\thresh}{4}
.\end{equation}

\begin{claim}\label{lm:error-fail}
$\Pr[\Error_{2}] \leq \frac{\delta}{2}$
\end{claim}
\begin{proof*}

We decompose  $\Pr[\Error_{2}]$ based on the value of $p$ at the end of Algorithm~\ref{algo:bfinal}. To this end, let us define the event
    $\bad$: ``The value of $p$ at line~\ref{line:boutput} in Algorithm~\ref{algo:bfinal} is less than  $\frac{\thresh}{4\Fo}$.''

Let $\ell =  \lfloor \log (\frac{4\Fo}{\thresh} ) \rfloor $. Since every value of $p$ can be expressed as power of $2$, we have that  $p < 2^{-\ell}$ if and only if $p < \frac{\thresh}{4\Fo}$. 
Observe that $\Pr[\Error_{2}] \leq  \Pr[\bad ] + \Pr [\Error_{2} \cap   \overline{\bad}].$ 
%We will upper bound $\Pr[\bad]$ and $\Pr [\Error_{2}  \cap  \overline{\bad}]$ separately. 

% For any $j \in [1, m]$ and $a\in S_j$ let $r_a$ denote the indicator random variable indicating whether $a$ is in the set $\mathcal{Y}_{\ell}$.  By Claim~\ref{cl:loop} the random variables $\{r_a\}_{a\in S_j}$ are independent and for all $a\in S_j$ $\Pr[r_a = 1] = 2^{\ell}$. 
% %
% Since, $|\mathsf{Y}_{\ell,j}| = \sum_{a\in S_{j}} r_a$
% \begin{equation}\label{eq:Yellj}\mathbb{E}\left[|\mathsf{Y}_{\ell,j}|\right] = \mathbb{E}\left[\sum_{a\in S_{j}}r_a\right] = \sum_{a\in S_{j}}\Pr[r_a = 1] = 2^{\ell}\cdot |S_{j}| \leq  2^{\ell} \cdot \Fo \leq \frac{\thresh}{4}.\end{equation}

\paragraph*{Bounding $\Pr[\bad] $}

For $j \in [1,m]$, let $\Bad{j}$ denote the event that `$j$th iteration of the \textbf{for} loop in Algorithm~\ref{algo:bfinal} is the first iteration where the value of $p$ goes below $2^{-\ell}$' i.e., $\p_{j-1} = 2^{-\ell}$ and $\p_{j} = 2^{-(\ell+1)}$.  
%Recall that in every iteration of the loop, the value of $p$ can decrease at most by a factor of $\frac{1}{2}$ and cannot increase. 
Therefore,  $\Pr[\bad] = \sum_{j=1}^{m} \Pr[\Bad{j}] $. We will now compute $\Pr[\Bad{j}]$ for a fixed $j$. 
%{\color{red} Observe that $\Pr[\Bad{j}] \leq \Pr[|\mathsf{Y}_{\ell,j}| \geq \thresh]$}
%

By definition of $\Bad{j}$, we have  $|\mathcal{X}| = \thresh$ and $p=2^{-\ell}$ in line~\ref{line:bIfbegin} of Algorithm~\ref{algo:bfinal} for $i=j$, i.e., $|\mathsf{Y}_{\ell, j}| = \thresh$. 
From Equation~\ref{eq:Yellj} we have $\mathbb{E}\left[|\mathsf{Y}_{\ell,j}|\right] \leq 2^{\ell} \cdot \Fo \leq \frac{\thresh}{4}.$
Thus, 
$\Pr[\Bad{j}] \leq \Pr[|\mathsf{Y}_{\ell,j}| \geq \thresh]
 \leq 2e^{-\frac{9 \cdot \thresh}{20}}   \leq \frac{\delta}{4m}$, where the second inequality follows from  Chernoff Bound. 
%\end{align*}
%
Therefore, $\Pr[\bad] \leq \frac{\delta}{4}$.

\paragraph*{Bounding $\Pr [\Error_{2}\cap \overline{\bad}]$}
%
%
%

%
%Similar to the above, for $a\in S_j$, let $r_a$ denote the indicator random variable indicating whether $a$ is in the set $\mathcal{Y}_{m,q}$. By Claim~\ref{cl:loop}, the random variables $ \{r_a\}_{a\in S_j} $ are independent and for all $a\in S_j$, we have $\Pr[r_a = 1] = q$. Thus $|\mathcal{Y}_{m,q}| = \sum_{a\in S_j}r_a$ and thus $\mathbb{E}[|\mathcal{Y}_{m,q}|] = q \cdot \Fo$. 
%
Let us define the event
$\Error_{2,q}:$ `$\p_m = 2^{-q}$ and $ \frac{|\mathsf{X}_{m}|}{2^{-q}} \notin [(1-\epsilon)\Fo, (1+\epsilon)\Fo]$ ' 

\smallskip

\noindent Observe that $\Pr[\Error_{2,q}] \leq \Pr[|\mathsf{Y}_{q,m}| \notin [(1-\epsilon)\cdot\frac{\Fo}{2^q}, (1+\epsilon)\cdot\frac{\Fo}{2^q}]]$. 
\vspace{-1em}
\begin{align*}
  \text{Therefore, }  \Pr [\Error_{2} \cap  \overline{\bad}] &\leq 
  %\Pr \left[\Error_{2}  \right] \leq 
  \sum_{q =0}^{\ell} \Pr[\Error_{2,q}] \\
    &\leq \sum_{q =0}^{\ell} \Pr[|\mathsf{Y}_{q,m}| \notin [(1-\epsilon)\cdot\frac{\Fo}{2^q}, (1+\epsilon)\cdot\frac{\Fo}{2^q}]]\\
 & \leq \sum_{q = 0}^{\ell} 2  e^{-\frac{\varepsilon^2 \Fo}{3 \cdot 2^q}} \hfill \ \ \ \ \ [\text{{Using Equation~\ref{eq:Yellj} and Chernoff bound}}]\\
%    &\leq  e^{-\frac{\varepsilon^2\thresh}{12}} + e^{-\frac{\varepsilon^2\cdot 2 \cdot \thresh}{12}} + \ldots  +  \hfill \ \ \ \ \ [\text{{Using Chernoff bound}}]\\
  %  & \leq \frac{e^{-\frac{\varepsilon^2\thresh}{12}}}{1-e^{-\frac{\varepsilon^2\thresh}{12}}} \leq  \frac{\frac{\delta}{8m}}{1-\frac{\delta}{8m}} \leq \frac{\delta}{8m-\delta} \leq \frac{\delta}{7} 
  & \leq 4 e^{-\frac{\varepsilon^2 \Fo}{3 \cdot 2^{\ell}}} \leq 4e^{-\frac{\varepsilon^2\thresh}{12}} \leq 4\cdot \left(\frac{\delta}{8}\right)^{\log e} \leq  \frac{\delta}{4} \qedhere 
                      \end{align*}

\end{proof*}

%& \leq \sum_{q = \ell}^0 e^{-frac{\varepsilon^2 2^q\Fo}{3}} \hfill \ \ \ \ \ [\text{{Using Equation~\ref{eq:Yellj} and Chernoff bound}}]\\

%

%
%
%
%

%
%
%

%
%

%

%

%

%

%

\section{Bibliographic Remarks}\label{sec:related}

Distinct Elements problem (or $\Fo$ estimation problem) is one of the most investigated  problem in the data streaming model ~\cite{AMS99,BJKST02,BKS02,Blasiok18,BC09,DF03,CVF03,FFGM07, FM85,GT01,IW03,KNW10}. While the Distinct Elements problem has a wide range of applications in several areas of computing, it was first investigated in the algorithms community by Flajolet and Martin~\cite{FM85}. They provided the first approximation under the assumption of the existence of hash functions with full independence. The seminal work of Alon, Matias, and Szegedy~\cite{AMS99} that introduced the data streaming model of computation revisited this problem as a special case of $F_k$ estimation problem and achieved space complexity of $O(\log n)$ for $\varepsilon > 1$ and constant $\delta$. The first $(\varepsilon,\delta)$ approximation for Distinct Elements problem was Gibbson and Tirthpura who achieved $O(\frac{\log n}{\varepsilon^2})$ space complexity~\cite{GT01}. Bar-Yossef, Jayram, Kumar, Sivakumar and Trevisan improved the space complexity bound to $\Tilde{O}(\log n + 1/\varepsilon^2)$~\cite{BJKST02}. Subsequently, Kane, Nelson, and Woodruff achieved $O(\log n + 1/\varepsilon^2)$ which is optimal in $n$ and $\varepsilon$~\cite{KNW10}. All the above bounds are for a fixed confidence parameter $\delta$, which can be amplified to achieve confidence bounds for arbitrary $\delta$ by simply running $\log (\frac{1}{\delta})$-estimators in parallel and returning the median. This incurs a multiplicative factor of $\log (\frac{1}{\delta})$. B\l{}asiok designed an $(\varepsilon,\delta)$ approximation algorithm for $\Fo$ estimation problem with space complexity of $O(\frac{1}{\varepsilon^2}\cdot\log \frac{1}{\delta} + \log n)$, thereby matching the lower bound in all the three parameters $n,\varepsilon$ and $\delta$~\cite{Blasiok18}. As is expected, every subsequent improvement added to the complexity of the algorithm or the analysis, and a majority of these work remain beyond the reach of non-experts. A crucial technical ingredient for all the works mentioned above is their careful usage of limited-independence hash functions in order to make space $\poly(\log n)$. Monte Carlo-based approaches have been utilized in the context of size estimation of the union of sets, but their straightforward adaptation to the streaming setting did not seem to yield progress. Recently, a new sampling-based approach was proposed in the context of estimating the size of the union of sets in the streaming model that achieves space complexity with $\log m$-dependence~\cite{MVC21}. The algorithm we presented adapts ideas from this work to the context of $\Fo$ estimation.

\section*{Acknowledgments}
We are deeply grateful to Donald E. Knuth for his thorough review, which not only enhanced the quality of this paper (including fixing several errors) but has also inspired us for higher standards. 
We owe sincere thanks to Arijit Ghosh, Mridul Nandi, Soumit Pal, Uddalok Sarkar, and anonymous reviewers for careful reading and pointing out critical errors in earlier versions. 
This work was supported in part by National Research Foundation Singapore under its NRF Fellowship Programme (NRF-NRFFAI1-2019-0004), Singapore Ministry of Education Academic Research Fund Tier 1, Ministry of Education Singapore Tier 2 grant (MOE-T2EP20121-0011),  and Amazon Faculty Research Awards.  Vinod was supported in part by NSF CCF-2130608 and NSF HDR:TRIPODS-1934884 awards. 


\begin{thebibliography}{10}

\bibitem{AMS99}
Noga Alon, Yossi Matias, and Mario Szegedy.
\newblock The space complexity of approximating the frequency moments.
\newblock {\em J. Comput. Syst. Sci.}, 58(1):137--147, 1999.
\newblock \href {https://doi.org/10.1006/jcss.1997.1545}
  {\path{doi:10.1006/jcss.1997.1545}}.

\bibitem{BJKST02}
Ziv Bar{-}Yossef, T.~S. Jayram, Ravi Kumar, D.~Sivakumar, and Luca Trevisan.
\newblock Counting distinct elements in a data stream.
\newblock In {\em Proc. of {RANDOM}}, pages 1--10, 2002.
\newblock \href {https://doi.org/10.1007/3-540-45726-7\_1}
  {\path{doi:10.1007/3-540-45726-7\_1}}.

\bibitem{BKS02}
Ziv Bar{-}Yossef, Ravi Kumar, and D.~Sivakumar.
\newblock Reductions in streaming algorithms, with an application to counting
  triangles in graphs.
\newblock In {\em Proceedings of SODA}, pages 623--632, 2002.
\newblock URL: \url{http://dl.acm.org/citation.cfm?id=545381.545464}.

\bibitem{Blasiok18}
Jaroslaw Blasiok.
\newblock Optimal streaming and tracking distinct elements with high
  probability.
\newblock In {\em Proc. of SODA}, 2018.
\newblock \href {https://doi.org/10.1137/1.9781611975031.156}
  {\path{doi:10.1137/1.9781611975031.156}}.

\bibitem{BC09}
Joshua Brody and Amit Chakrabarti.
\newblock A multi-round communication lower bound for gap hamming and some
  consequences.
\newblock In {\em Proc. of CCC}, pages 358--368. {IEEE} Computer Society, 2009.
\newblock \href {https://doi.org/10.1109/CCC.2009.31}
  {\path{doi:10.1109/CCC.2009.31}}.

\bibitem{DF03}
Marianne Durand and Philippe Flajolet.
\newblock Loglog counting of large cardinalities (extended abstract).
\newblock In {\em Proc. of ESA},
   pages 605--617, 2003.
\newblock \href {https://doi.org/10.1007/978-3-540-39658-1\_55}
  {\path{doi:10.1007/978-3-540-39658-1\_55}}.

\bibitem{CVF03}
Cristian Estan, George Varghese, and Michael~E. Fisk.
\newblock Bitmap algorithms for counting active flows on high-speed links.
\newblock {\em {IEEE/ACM} Trans. Netw.}, 14(5):925--937, 2006.
\newblock \href {https://doi.org/10.1145/1217709} {\path{doi:10.1145/1217709}}.

\bibitem{FFGM07}
Philippe Flajolet, {\'E}ric Fusy, Olivier Gandouet, and Fr{\'e}d{\'e}ric
  Meunier.
\newblock Hyperloglog: the analysis of a near-optimal cardinality estimation
  algorithm.
\newblock In {\em Discrete Mathematics and Theoretical Computer Science}, pages
  137--156, 2007.
\newblock URL: \url{https://doi.org/10.46298/dmtcs.3545}.

\bibitem{FM85}
Philippe Flajolet and G.~Nigel Martin.
\newblock Probabilistic counting algorithms for data base applications.
\newblock {\em J. Comput. Syst. Sci.}, 31(2):182--209, 1985.
\newblock \href {https://doi.org/10.1016/0022-0000(85)90041-8}
  {\path{doi:10.1016/0022-0000(85)90041-8}}.

\bibitem{GT01}
Phillip~B. Gibbons and Srikanta Tirthapura.
\newblock Estimating simple functions on the union of data streams.
\newblock In {\em Proc. of {SPAA}}, pages 281--291, 2001.
\newblock \href {https://doi.org/10.1145/378580.378687}
  {\path{doi:10.1145/378580.378687}}.

\bibitem{IW03}
Piotr Indyk and David~P. Woodruff.
\newblock Tight lower bounds for the distinct elements problem.
\newblock In {\em Proc. of FOCS}, pages 283--288, 2003.
\newblock \href {https://doi.org/10.1109/SFCS.2003.1238202}
  {\path{doi:10.1109/SFCS.2003.1238202}}.

\bibitem{KNW10}
Daniel~M. Kane, Jelani Nelson, and David~P. Woodruff.
\newblock An optimal algorithm for the distinct elements problem.
\newblock In {\em Proc. of {PODS}}, pages 41--52, 2010.
\newblock \href {https://doi.org/10.1145/1807085.1807094}
  {\path{doi:10.1145/1807085.1807094}}.

\bibitem{MVC21}
Kuldeep~S. Meel, N.~V. Vinodchandran, and Sourav Chakraborty.
\newblock Estimating the size of union of sets in streaming models.
\newblock In {\em Proc. of PODS}, pages 126--137, 2021.
\newblock \href {https://doi.org/10.1145/3452021.3458333}
  {\path{doi:10.1145/3452021.3458333}}.

\end{thebibliography}
\end{document}